\documentclass[epj]{svjour}
\usepackage{graphics}
\usepackage{amsmath}
\usepackage{amssymb}
\usepackage{morefloats}
\usepackage[usenames,dvipsnames]{xcolor}
\usepackage{blkarray}
\usepackage{verbatim}
\usepackage{hyperref}
\usepackage{color}

\begin{document}
\title{Statistical Physics of the Spatial Prisoner's Dilemma with Memory-Aware Agents}
\author{Marco Alberto Javarone\inst{1,2}
}                     
%
%
\institute{Department of Mathematics and Computer Science, University of Cagliari, Cagliari (Italy) \and DUMAS - Department of Humanities and Social Sciences, University of Sassari, Sassari (Italy)}
\date{Received: date / Revised version: date}
%
\abstract{
We introduce an analytical model to study the evolution towards equilibrium in spatial games, with `memory-aware' agents, i.e., agents that accumulate their payoff over time. In particular, we focus our attention on the spatial Prisoner's Dilemma, as it constitutes an emblematic example of a game whose Nash equilibrium is defection.
Previous investigations showed that, under opportune conditions, it is possible to reach, in the evolutionary Prisoner's Dilemma, an equilibrium of cooperation. Notably, it seems that mechanisms like motion may lead a population to become cooperative.
In the proposed model, we map agents to particles of a gas so that, on varying the system temperature, they randomly move.
In doing so, we are able to identify a relation between the temperature and the final equilibrium of the population, explaining how it is possible to break the classical Nash equilibrium in the spatial Prisoner's Dilemma when considering agents able to increase their payoff over time.
Moreover, we introduce a formalism to study order-disorder phase transitions in these dynamics.
As result, we highlight that the proposed model allows to explain analytically how a population, whose interactions are based on the Prisoner's Dilemma, can reach an equilibrium far from the expected one; opening also the way to define a direct link between evolutionary game theory and statistical physics.
} 
\maketitle
\section{Introduction}\label{intro}
Evolutionary games~\cite{perc03,nowak04,tomassini01} represent the attempt to study the evolution of populations~\cite{moreno02,moreno03,hofbauer01} by the framework of game theory~\cite{dilemma01}. Notably, these games allow to analyze simplified scenarios in different domains, spanning from socio-economic dynamics to biological systems~\cite{perc01,perc02,perc03,perc04,perc05,perc06,friedman01,shuster01,frey01,nowak01,nowak02}.
In general, evolutionary games consider a population of agents whose interactions are based on games like the Prisoner's Dilemma (hereinafter PD) or the Hawk-Dove game~\cite{moreno02}, where there are two possible strategies: cooperation and defection. As in classical game theory, the concept of equilibrium represents a core aspect~\cite{galam01}. Therefore, we aim to evaluate if a population reaches an equilibrium equal or different from the expected one, i.e., the Nash equilibrium of the considered game.
At each interaction, agents gain a payoff according to the adopted strategy and to a payoff matrix.
The payoff represents a form of reward in the considered domain (e.g., money in an economic system or food in an ecosystem).
Remarkably, as agents are allowed to change their strategy over time, we can map them to spins with states $\sigma = \pm 1$, representing cooperation and defection, respectively. In doing so, we can analyze order-disorder transitions in the spatial PD.
Previous studies~\cite{meloni01,tomassini02,tomassini03,tomassini04,moreno01,javarone01,javarone02} have shown that, under particular conditions, it is possible that a population playing a game like the PD, i.e., a game characterized by defection as Nash equilibrium, can be able to reach a final state of full cooperation.
For instance, it seems that both motion~\cite{meloni01,tomassini02,tomassini03,tomassini04} and competitiveness~\cite{javarone01} can lead an agent population to cooperate~\cite{nowak03} and, more in general, spatial structure plays a key role in the evolution of cooperation~\cite{szabo01,nowak05}.
Usually, adding properties to agents, as motion, conformity and competitiveness, entails to increase the complexity of the resulting model. Thus, most investigations on evolutionary games are based on computational approaches.
Therefore, in this work we try to provide an analytical description of the spatial PD, in order to explain how a population can become cooperative and to strengthen the link between evolutionary game theory~\cite{szabo02} and statistical physics~\cite{huang01}. It is worth to highlight that we consider `memory-aware' agents, i.e., agents that accumulate their payoff over time. Remarkably, this last condition represents the major difference with most of the evolutionary game models studied by computational approaches (see for instance~\cite{perc07,perc08}). On the other hand, considering `memory-aware' agents makes the problem more tractable from an analytical perspective.
The remainder of the paper is organized as follows: Section~\ref{sec:model} introduces the proposed model and its analytical formulation. Section~\ref{sec:results} shows analytical results. Eventually, Section~\ref{sec:conclusions} ends the paper.
\section{Model}\label{sec:model}
In the proposed model, we are interested in studying the spatial prisoner's dilemma by an analytical approach. Let us start by introducing the general form of a payoff matrix
\begin{equation}\label{eq:payoff}
\bordermatrix{~ & C & D \cr
                  C & R & S \cr
                  D & T & P \cr}
\end{equation}
\noindent where the set of strategies is $\Sigma = \left\{C,D\right\}$: \textit{C} stands for `Cooperator' and \textit{D} for `Defector'. 
In the matrix~\ref{eq:payoff}, \textit{R} is the gain obtained by two interacting cooperators, \textit{T} represents the \textit{Temptation}, i.e., the payoff that an agent gains if it defects while its opponent cooperates, $S$ the \textit{Sucker's payoff}, i.e., the gain achieved by a cooperator while the opponent defects, eventually $P$ the payoff of two interacting defectors.
In the case of the PD, matrix elements of~\ref{eq:payoff} are: $R = 1$, $0 \le S \le -1$, $1 \le T \le 2$ and $P = 0$.
As stated before, during the evolution of the system agents can change their strategy from $C$ to $D$, and vice versa, following an updating rule, as for instance the one named `imitation of the best' (see~\cite{meloni01,moreno02}), where agents imitate the strategy of their richest neighbor.
\subsection{Mean field approach}
Now, we consider a mixed population of $N$ agents with, at the beginning, an equal density of cooperators and defectors. 
Under the hypothesis that all agents interact together, at each time step the payoffs gained by cooperators and defectors are computed as follows
\begin{equation}\label{eq:payoff_total}
\begin{cases}
\pi_c = (\rho_c \cdot N -1) + (\rho_d \cdot N) S\\
\pi_d = (\rho_c \cdot N) T
\end{cases}
\end{equation}
\noindent with $\rho_c + \rho_d = 1$, $\rho_c$ density of cooperators and $\rho_d$ density of defectors. We recall that defection is the dominant strategy in the PD and, even if we set $S = 0$ and $T = 1$, it corresponds to the final equilibrium because $\pi_d$ is always greater than $\pi_c$.
At this point, it is important to highlight that previous investigations~\cite{meloni01,tomassini02,tomassini03} have been performed by `memoryless' agents (i.e., agents that do not accumulate the payoff over time) whose interactions were defined only with their neighbors, and focusing only on one agent (and on its neighbors) at a time.
These conditions are fundamental. For instance, if at each time step we randomly select one agent interacting only with its neighbors, there exists the probability to select consecutively a number of close cooperators; thus, in this occurrence, very rich cooperators may emerge and then prevail on defectors, even without introducing mechanisms like motion.
It is also worth to observe that as $P = 0$, a homogeneous population of defectors does not increase its overall payoff. Instead, according to the matrix~\ref{eq:payoff}, a cooperative population continuously increases its payoff over time. 
\\
Now, we consider a population divided into two groups by a wall: a group $G^a$ composed of cooperators, and a mixed group $G^b$, i.e., composed of cooperators and defectors in equal amount.
Agents interact only with members of the same group, then the group $G^a$ never changes and, in addition, it strongly increases its payoff over time. The opposite occurs in the group $G^b$, as it converges to an ordered phase of defection, limiting its final payoff.
Remarkably, in this scenario, we can introduce a strategy to modify the equilibria of the two groups. In particular, we can both change to cooperation the equilibrium of $G^b$, and to defection that of $G^a$. In the first case, we have to wait a while, before moving one or few cooperators to $G^b$, so that defectors increase their payoff, but during the revision phase they change strategy to cooperation as the newcomers are richer than them. In the second case, if we move after few time steps a small group of defectors from $G^b$ to $G^a$, the latter converges to a final defection phase.
These preliminary and theoretical observations let emerge an important property of the `memory-aware' PD: considering the two different groups, cooperators may succeed when act after a long time and individually. Instead, defectors may succeed acting fast and in group. Notably, rich cooperators have to move individually since otherwise many rich cooperators risk to increase too much the payoff of defectors that, in this case, will not change strategy. The opposite holds for defectors that, acting in group, may strongly reduce the payoff of a community of cooperators (for $S < 0$).
\subsubsection{Mapping agents to gas particles}
We hypothesize that the spatial PD, with moving agents, can be successfully studied by the framework of kinetic theory~\cite{huang01}. 
Therefore, in the proposed model, we map agents to particles of a gas. In doing so, the average speed of particles is computed as $<v> = \sqrt{\frac{3 T_s k_b}{m_p}}$, with $T_s$ system temperature, $k_b$ Boltzmann constant, and $m_p$ particle mass.
Particles are divided into two groups by a permeable wall, so that it can be crossed by particles, but it avoids interactions among particles belonging to different groups.
Now, it is worth to emphasize that we can provide a dual description of our system: one in the `physical' domain of particles, the other in the `information' domain of agents. Notably, to analyze the system in the `information' domain we will introduce, as above discussed, the mapping of agents to a spin system (see~\cite{javarone03}).
Summarizing, we map agents to gas particles in order to represent their `physical' property of motion, and we map agents to spins for representing their `information' property (i.e., their strategy). Remarkably, these two mappings can be viewed as two different layers for studying how the agent population evolves over time. Although the physical property (i.e., the motion) affects the agent strategy (i.e., its spin), the equilibrium can be reached in both layers/domains independently. This last observation is important since we are interested in evaluating only the final equilibrium reached in the 'information' domain. 
Then, as stated before, agents interact only with those belonging to the same group, so the evolution of the mixed group $G^b$ can be described by following equations
\begin{equation}\label{eq:lotka_volterra}
\begin{cases}\frac{d\rho_c^b(t)}{dt} = p_c^b(t) \cdot \rho_c^b(t) \cdot \rho_d^b(t) - p_d^b(t) \cdot \rho_d^b(t) \cdot \rho_c^b(t)\\ 
\frac{d\rho_d^b(t)}{dt} =  p_d^b(t)  \cdot \rho_d^b(t) \cdot \rho_c^b(t) - p_c^b(t) \cdot \rho_c^b(t) \cdot \rho_d^b(t)\\
\rho_c^b(t) + \rho_d^b(t) = 1
\end{cases}
\end{equation}
\noindent with $p_c^b(t)$ probability that cooperators prevail on defectors (at time $t$), and $p_d^b(t)$ probability that defectors prevail on cooperators (at time $t$). These probabilities are computed according to the payoffs obtained, at each time step, by cooperators and defectors
\begin{equation}\label{eq:prob_payoff}
\begin{cases} p_c^b(t) = \frac{\pi_c^{b}(t)}{\pi_c^{b}(t) + \pi_d^{b}(t)}\\ 
p_d^b(t) = 1 - p_c^b(t)
\end{cases}
\end{equation}
The system~\ref{eq:lotka_volterra} can be analytically solved provided that, at each time step, values of $p_c^b(t)$ and $p_d^b(t)$ be updated. So, the density of cooperators reads
\begin{equation}\label{eq:density_cooperators}
\rho_c^b(t) = \frac{\rho_c^b(0)}{\rho_c^b(0) - [(\rho_c^b(0) - 1)\cdot e^{\frac{\tau t}{N^b}}]}
\end{equation}
\noindent with $\rho_c^b(0)$ initial density of cooperators in $G^b$, $\tau = p_d^b(t) - p_c^b(t)$, and $N^b$ number of agents in $G^b$.
Recall that setting $T_s = 0$, not allowed in a thermodynamic system, corresponds to a motionless case, leading to the Nash equilibrium in $G^b$. Instead, for $T_s > 0$ we can find more interesting scenarios. %
Now we suppose that, at time $t = 0$, particles of $G^a$ are much closer to the wall than those of $G^b$ (later we will relax this constraint); for instance, let us consider a particle of $G^a$ that, during its random motion, it is following a trajectory of length $d$ (in the $n$-dimensional physical space) towards the wall. Assuming this particle is moving with speed equal to $<v>$, we can compute the instant of crossing $t_c = \frac{d}{<v>}$, i.e., the instant when it moves from $G^a$ to $G^b$.
Thus, on varying the temperature $T_s$, we can vary $t_c$. \\
Let us consider the payoff of cooperators in the two groups. Each cooperator in $G^a$ gains
\begin{equation}\label{eq:payoff_g_a}
\pi_c^a =  (\rho_c^a \cdot N^a - 1) \cdot t
\end{equation}
On the other hand, the situation for cooperators in $G^b$ is much more different as, according to the Nash equilibrium, their amount decreases over time. Therefore, we can consider how changes the payoff of the last cooperator survived in $G^b$ 
\begin{equation}\label{eq:payoff_g_b}
\pi_c^{b} = \sum_{i = 0}^{t}[(\rho_c^b \cdot N^b - 1) + (\rho_d^b \cdot N^b) S]_i
\end{equation}
\noindent moreover, $\pi_c^{b} \to 0$ as $\rho_c^{b} \to 0$.
At $t = t_c$, a new cooperator reaches $G^b$, with a payoff computed with equation~\ref{eq:payoff_g_a}.
\section{Results}\label{sec:results}
\begin{figure*}[t]
\resizebox{1.0\textwidth}{!}{
\includegraphics{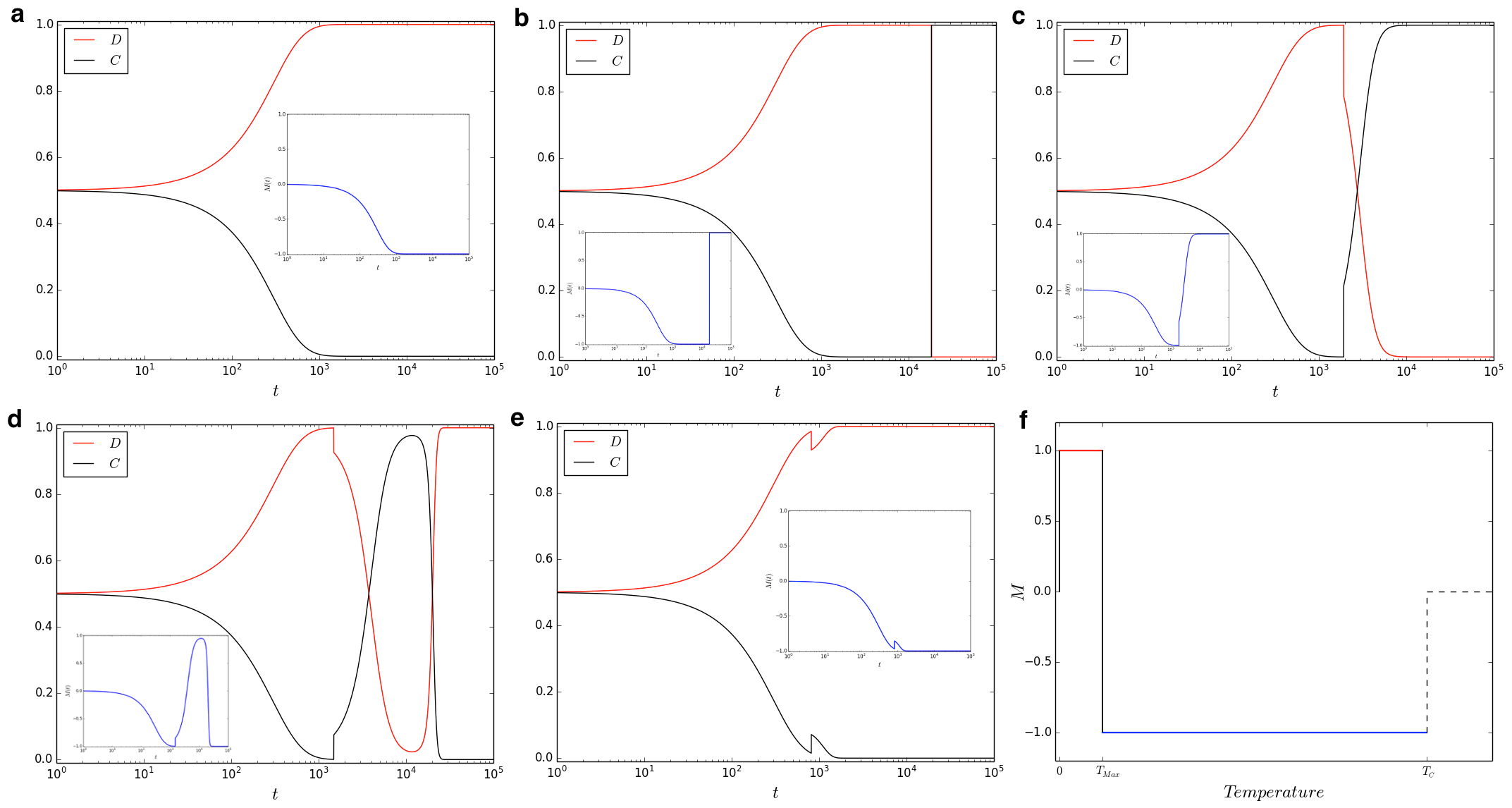}}
\caption{\small From \textbf{a} to \textbf{e}: Evolution of the group $G^b$, with $N = 100$ and $\epsilon = 1$, on varying the temperature: \textbf{a.} $T_s = 0$. \textbf{b.} $T_s = 0.1$. \textbf{c.} $T_s = 9$. \textbf{d.} $T_s = 15$. \textbf{e.} $T_s = 50$. Insets show the system magnetization over time. The istant $t = t_c$, can be detected in plots \textbf{c,d,e} as a discontinuity of the two lines (i.e., red and black). \textbf{f.} Final magnetization $M$, of $G^b$, for different temperatures ($T_c$ indicates the `critical temperature'). \label{fig:evolution}}
\end{figure*}

\begin{figure}[!h]
\resizebox{0.49\textwidth}{!}{
\includegraphics{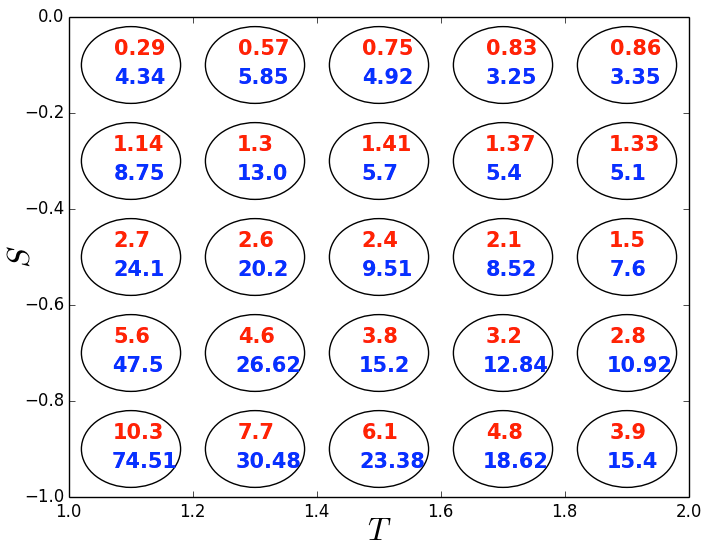}}
\caption{\small Maximum values of temperature $T_s$ that allow the group $G^b$ to converge to cooperation. Red values correspond to results computed with $\epsilon = 0.5$, while blue values to those computed with $\epsilon= 1$. Circles are placed in the $TS$ diagram indicating values of $T$ and $S$, of the payoff matrix, used for each case. Even for high values of $T$, and small values of $S$, it is possible to achieve cooperation. \label{fig:st_plot}}
\end{figure}

\begin{figure}[!h]
\resizebox{0.49\textwidth}{!}{
\includegraphics{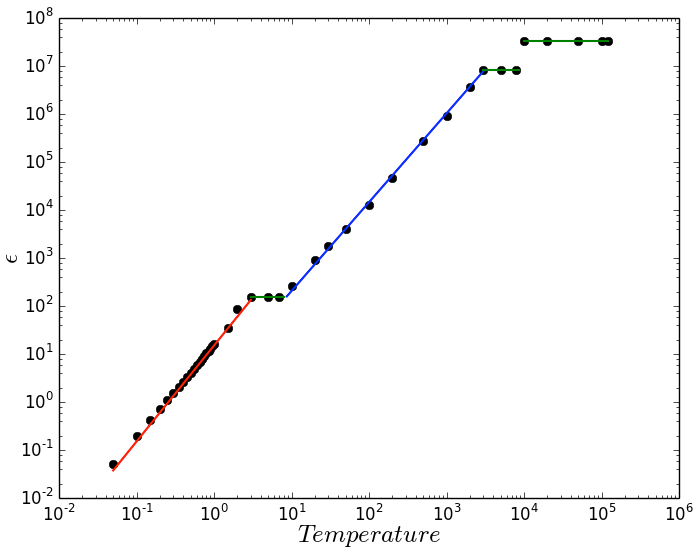}}
\caption{\small Maximum value of system temperature that allows to achieve cooperation at equilibrium versus $\epsilon$ (i.e., the ratio between particles in the two groups). Different colors identify different trends, fitted by power-law functions. After the final green plateau, temperatures are too high to play the spatial PD. \label{fig:epsilon_plot}}
\end{figure}

\begin{figure}[!h]
\resizebox{0.49\textwidth}{!}{
\includegraphics{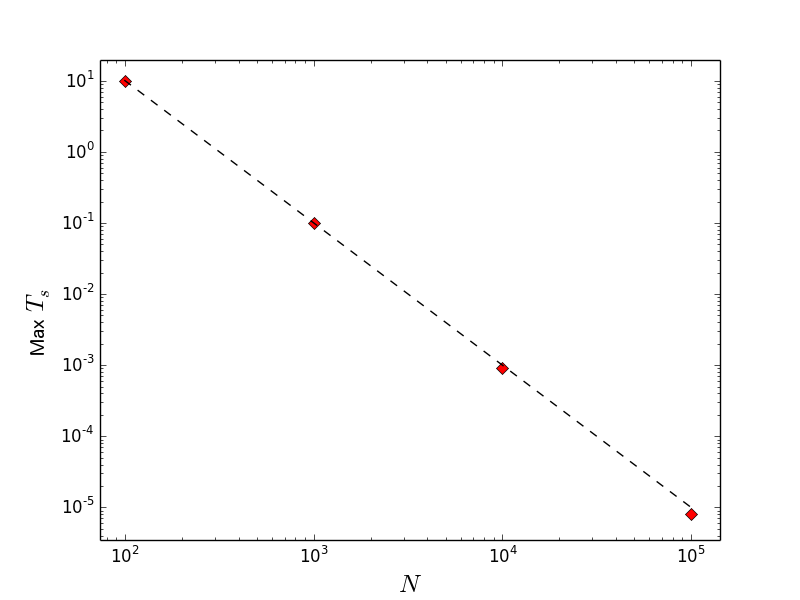}}
\caption{\small Maximum value of $T_s$ to achieve full cooperation at equilibrium in function of $N$, i.e., the size of the population. The fitting function (dotted line) is a power-law characterized by a scaling parameter equal to $2$. \label{fig:scaling}}
\end{figure}

The analytical solution~\ref{eq:density_cooperators} allows to analyze the evolution of the system and to evaluate how initial conditions affects the outcomes of the model.
Let us observe that, if $\pi_c^a(t_c)$ is enough big, the new cooperator may modify the equilibrium of $G^b$, turning defectors to cooperators. Notably, the payoff considered to compute $p_c^{b}$, after $t_c$, corresponds to $\pi_c^a(t_c)$, as the newcomer is the richest cooperator in $G^b$.
Furthermore, we note that $\pi_c^a(t_c)$ depends on $N^a$, hence we study the evolution of the system on varying the parameter $\epsilon = \frac{N^a}{N^b}$, i.e., the ratio between particles in the two groups. Eventually, for numerical convenience, we set $k_b = 1\cdot10^{-8}$, $m_p = 1$, and $d = 1$.
\\
Figure~\ref{fig:evolution} shows the evolution of $G^b$, for $\epsilon = 1$ on varying $T_s$ and, depicted in the inner insets, the variation of system magnetization over time (always inside $G^b$) computed as~\cite{mobilia01}
\begin{equation}\label{eq:magnetization}
M = \frac{\sum_{i=1}^{N^b} \sigma_i}{N^b}
\end{equation}
\noindent with $\sigma_i$ strategy of the $i$-agent.
As discussed before, in the physical domain of particles, heating the system entails the average speed of particles increases. Thus, under the assumption that two agents play together if they stay close (i.e., in the same group) for a long enough time, we hypothesize that exists a maximum speed such that for greater values interactions do not occur (in terms of game). This hypothesis requires a critical temperature $T_c$, above which no interactions, in the `information' domain, are possible. 
As shown in plot \textbf{f} of figure~\ref{fig:evolution}, for temperatures in range $0 < T_s < T_{max}$ the system converges to a cooperation phase (i.e., $M = +1$), for $T_{max} < T_s < T_c$ the system follows the Nash equilibrium (i.e., $M = -1$), and for $T > T_c$ a disordered phase emerges at equilibrium.
Remarkably, results of our model suggest that it is always possible to compute a range of temperatures to obtain an equilibrium of full cooperation ---see figure~\ref{fig:st_plot}.
Moreover, we study the variation of $T_{max}$ on varying $\epsilon$ (see figure~\ref{fig:epsilon_plot}) showing that, even for low $\epsilon$, it is possible to obtain a time $t_c$ that allows the system to converge towards cooperation.
Eventually, we investigate the relation between the maximum value of $T_s$ that allows a population to become cooperative and its size $N$ (i.e., the number of agents).
Remarkably, as shown in figure~\ref{fig:scaling}, the maximum $T_s$ scales with $N$ following a power-law function characterized by a scaling parameter (i.e., an exponent) $\gamma \sim 2$. 
The value of $\gamma$ has been computed by considering values of $T_s$ shown in figure~\ref{fig:st_plot} for the case $\epsilon = 2$.
Eventually, it is worth to highlight that all analytical results let emerge a link between the system temperature and its final equilibrium. Recalling that we are not considering the equilibrium of the gas, i.e., it does not thermalize in the proposed model, we emphasize that the equilibrium is considered only in the information domain.
\subsection{Phase Transitions in the spatial PD}
\begin{figure*}[!t]
\resizebox{0.85\textwidth}{!}{
\includegraphics{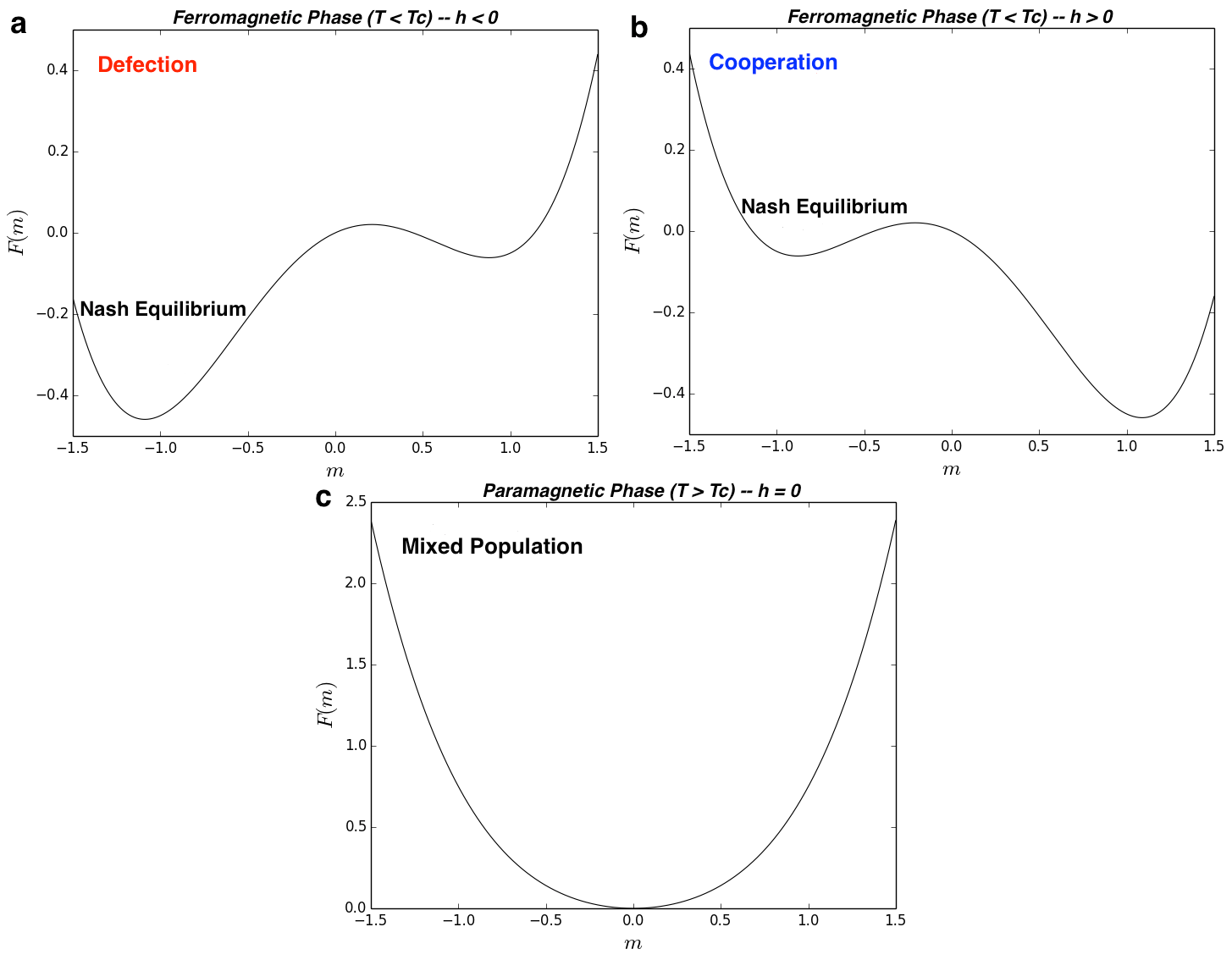}}
\caption{\small Order-disorder phase transitions in the population. For $T_s < T_c$, the population is in a ferromagnetic phase: \textbf{a.} Applying an external negative field, the system converges to the Nash equilibrium, corresponding to $m = -1$ (as $\sigma = -1$ represents defection); \textbf{b.} Applying an external positive field, the population converges to cooperation ($\sigma = +1$), corresponding to $m = +1$. \textbf{c.} For temperatures higher than $T_c$, a disordered paramagnetic phase emerges.\label{fig:landau}}
\end{figure*}

As discussed before, in the information domain we can study the system by mapping agents to spins, whose value represents their strategy. In addition, we can map the difference between winning probabilities, of cooperators and defectors, to an external magnetic field: $h = p_c^b - p_d^b$. In doing so, by the Landau theory~\cite{huang01}, we can analytically identify an order-disorder phase transition.
Notably, we analyze the free energy $F$ of the spin system on varying the control parameter $m$~\cite{barra01} (corresponding to the magnetization $M$)
\begin{equation}\label{eq:landau}
F(m) = -hm \pm \frac{m^2}{2} + \frac{m^4}{4}
\end{equation}
\noindent where the sign of the second term depends on the temperature, i.e., positive for $T_s > T_c$ and negative for $T_s < T_c$; recalling that $T_c$ represents the temperature beyond which it is not possible to play the PD due to the high particles speed (according to the condition before discussed). 
For the sake of clarity, we want to emphasize that the free energy is introduced in order to evaluate the nature of the final equilibrium achieved by the system. In particular, looking for the minima of $F$ allows to investigate if our population reaches the Nash equilibrium, or different configurations (e.g., full cooperation). Figure~\ref{fig:landau} shows a pictorial representation of the phase transitions that occur in our system, on varying $T_s$ and the external field $h$.
Finally, the constraints related to the average speed of particles, and to the distance between each group and the permeable wall, can in principle be relaxed as we can imagine to extend this description to a wider system with several groups (as done in previous investigations, e.g.~\cite{tomassini02}), where agents are uniformly spread in the whole space. 
It is worth to highlight that our results are completely in agreement with those achieved by authors who studied the role of motion in the PD (as~\cite{meloni01,tomassini02}), explaining why clusters of cooperators emerge in their simulations~\cite{tomassini02}. We also recall that, in the proposed model, we are using memory-aware agents, while in previous computational investigations agents reset their payoff at each step, i.e., before to start new interactions.
\section{Conclusions}\label{sec:conclusions}
To conclude, in this work we provide an analytical description of the spatial Prisoner's Dilemma, by using the framework of statistical physics, studying the particular case of agents provided with memory of their payoff (defined memory-aware agents). This condition entails that their payoff is not reset at each time step, so that they can increase it over time. 
In particular, we propose a model based on the kinetic theory of gases, showing how motion may lead a population towards an equilibrium far from the expected one (i.e., the Nash equilibrium). Remarkably, the final equilibrium depends on the system temperature, so that we have been able to identify a range of temperatures that triggers cooperation for all values of the payoff matrix (related to the PD). In addition, we found an interesting relation between the maximum temperature that foster cooperation and the size of the system. Notably, a scaling parameter in that relation has been computed by investigating different orders of magnitude of the size of the system.
Furthermore, the dynamics of the resulting model have been also described in terms of order-disorder phase transitions. Finally, we deem that our results open the way to define a direct link between evolutionary game theory and statistical physics.
%
%
%
%

\section*{Acknowledgments}
MAJ is extremely grateful to Adriano Barra for all priceless suggestions. Moreover, he wants to thank Mirko Degli Esposti, Marco Lenci, and Giampaolo Cristadoro for the useful comments. This work has been supported by Fondazione Banco di Sardegna.

\end{document}